\documentclass[fleqn]{annalen}
\usepackage{graphics}
\begin{document}
%%%%%%%%%%%%%%%%%%%%%%%%%%%%%%%%%%%%%%%%%%%%%%%%%%%%%%%%%%%%%%%%%%%%%%%%%%%%%%
%%%%%%%% the following newcommands will be completed by the publisher %%%%%%%%
%%%%%%%%%%%%%%%%%%%%%%%%%%%%%%%%%%%%%%%%%%%%%%%%%%%%%%%%%%%%%%%%%%%%%%%%%%%%%%
\newcommand{\volume}{xx}              %sets current volume,
\newcommand{\xyear}{1999}            %sets year in header
\newcommand{\issue}{xx}             %sets current issue,
\newcommand{\recdate}{dd.mm.yyyy}   %sets received date,
\newcommand{\revdate}{dd.mm.yyyy}    %sets revised date,
\newcommand{\revnum}{0}              %number of revisions,
\newcommand{\accdate}{dd.mm.yyyy}  %sets accepted date,
\newcommand{\coeditor}{xx}           %sets (co)editor,
\newcommand{\firstpage}{1}           %first page number,
\newcommand{\lastpage}{9}            %last page number,
\setcounter{page}{\firstpage}        %sets page counter to first page number
%%%%%%%%%%%%%%%%%%%%%%%%%%%%%%%%%%%%%%%%%%%%%%%%%%%%%%%%%%%%%%%%%%%%%%%%%%%%%%
%%%%%%%%%%%%%%%%%%%%%%%%%%%%%%%%%%%%%%%%%%%%%%%%%%%%%%%%%%%%%%%%%%%%%%%%%%%%%%
%%%%%%%%%%%%%%%%%% please give up to three keywords here %%%%%%%%%%%%%%%%%%%%%
%%%%%%%%%%%%%%%%%%%%%%%%%%%%%%%%%%%%%%%%%%%%%%%%%%%%%%%%%%%%%%%%%%%%%%%%%%%%%%
\newcommand{\keywords}{
Universal velocity correlations, mesoscopic systems, microwave billiards}
%%%%%%%%%%%%%%%%%%%%%%%%%%%%%%%%%%%%%%%%%%%%%%%%%%%%%%%%%%%%%%%%%%%%%%%%%%%%%%
%%%%%%%%%%%%%%%% please give up to three PACS numbers here %%%%%%%%%%%%%%%%%%%
%%%%%%%%%%%%%%%%%%%%%%%%%%%%%%%%%%%%%%%%%%%%%%%%%%%%%%%%%%%%%%%%%%%%%%%%%%%%%%
\newcommand{\PACS}{05.45.-a, 73.23.-b}
%%%%%%%%%%%%%%%%%%%%%%%%%%%%%%%%%%%%%%%%%%%%%%%%%%%%%%%%%%%%%%%%%%%%%%%%%%%%%%
%% please enter (First) Author (et al.) and short version of the title here %%
%%%%%%%%%%%% must not exceed 80 characters in length together %%%%%%%%%%%%%%%%
%%%%%%%%%%%%%%%%%%%%%%%%%%%%%%%%%%%%%%%%%%%%%%%%%%%%%%%%%%%%%%%%%%%%%%%%%%%%%%
\newcommand{\shorttitle}{M. Barth et al.,
Global and local level dynamics} %% sets the header on oddpage
%%%%%%%%%%%%%%%%%%%%%%%%%%%%%%%%%%%%%%%%%%%%%%%%%%%%%%%%%%%%%%%%%%%%%%%%%%%%%%
%%%%%%%%%%%%%%%%%%%%%%%% here comes the title group %%%%%%%%%%%%%%%%%%%%%%%%%%
%%%%%%%%%%%%%%%%%%%%%%%%%%%%%%%%%%%%%%%%%%%%%%%%%%%%%%%%%%%%%%%%%%%%%%%%%%%%%%
\title{Global and local level dynamics in chaotic microwave billiards}
%%%%%%%%%%%%%%%%%%%%%%%%%%%%%%%%%%%%%%%%%%%%%%%%%%%%%%%%%%%%%%%%%%%%%%%%%%%%%%
\author{M.\ Barth, U.\ Kuhl, and H.-J.\ St\"ockmann}
%%%%%%%%%%%%%%%%%%%%%%%%%%%%%%%%%%%%%%%%%%%%%%%%%%%%%%%%%%%%%%%%%%%%%%%%%%%%%%
\newcommand{\address}
  {Fachbereich Physik, Philipps-Universit\"at Marburg, Renthof 5,
  D-35032 Marburg, Germany}
%%%%%%%%%%%%%%%%%%%%%%%%%%%%%%%%%%%%%%%%%%%%%%%%%%%%%%%%%%%%%%%%%%%%%%%%%%%%%%
\newcommand{\email}{\tt michael.barth@physik.uni-marburg.de}
\maketitle
%%%%%%%%%%%%%%%%%%%%%%%%%%%%%%%%%%%%%%%%%%%%%%%%%%%%%%%%%%%%%%%%%%%%%%%%%%%%%
\begin{abstract}
The spectra of Sinai microwave billiards and rectangular billiards with
statistically distributed circular scatterers have been taken as a function
of the position of one wall, and of one of the scatterers, respectively.
Whereas in the first case the velocity distribution and correlations obey
the universal behaviour predicted by Simons and Altshuler, in the second
case a completely different behaviour is observed. This is due to the fact
that a shift of one wall changes the wave function globally whereas the
displacement of one scatterer only leads to a local perturbance.
\end{abstract}
%%%%%%%%%%%%%%%%%%%%%%%%%%%%%%%%%%%%%%%%%%%%%%%%%%%%%%%%%%%%%%%%%%%%%%%%%%%%%

\section{Introduction}

The eigenvalue dynamics of the spectra of disordered systems in dependence of
an external parameter has attracted considerable interest in recent years. It
has been motivated by the fact that the conductance through a system is
closely linked to the sensitivity of its eigenvalues on a perturbation
\cite{edw72,tho74}. In this respect the work of Akkermans and Montambaux has
to be mentioned in particular \cite{akk92}, who showed that the conductance
can be expressed in terms of the quadratically averaged eigenvalue
velocities,
\begin{equation}
C \sim \left< \left|\frac{\partial E_n}{\partial X}\right|^2 \right>.
\label{EqDefAC}
\end{equation}

Here the brackets denote a local average, and $X$ is the external parameter.
Later it was shown by Simons and Altshuler using supersymmetry techniques
that the eigenvalue velocities $v$ should be Gaussian distributed
\cite{sim93b}
\begin{equation}
P(v) = \frac{1}{\sqrt{2\pi\left<v^2\right>}}
\exp\left(-\frac{v^2}{2\left<v^2\right>}\right).
\label{EqGauss}
\end{equation}

Moreover, if the eigenvalues $E_n$ and the level dynamics parameter $X$ are
rescaled according to
\begin{equation}
\epsilon_n = \frac{E_n}{\Delta},
\qquad
x = \frac{1}{\Delta}
\sqrt{\left<\left|\frac{\partial E_n}{\partial X}\right|^2\right>} \: X,
\label{EqScale}
\end{equation}
where $\Delta$ is the mean level spacing, the velocity autocorrelation
function
\begin{equation}
c(x) = \left<
\frac{\partial\epsilon_n(x_0 + x)}{\partial x_0}
\cdot
\frac{\partial\epsilon_n(x_0)}{\partial x_0}
\right>
- \left< \frac{\partial\epsilon_n(x_0)}{\partial x_0} \right>^2
\label{EqGlobalAC}
\end{equation}
should obey a universal behaviour \cite{sza93,sim93b}. A recent review on
these questions can be found in \cite{sim95}.

There are a number of attempts to study the autocorrelation function
(\ref{EqGlobalAC}), both theoretically and experimentally. The systems under
consideration ranged from the hydrogen atom in a strong magnetic field
\cite{sim93c} via conformally deformed \cite{bru96} and ray-splitting
billiards \cite{hlu} to the acoustic spectra of vibrating quartz blocks
\cite{ber}. In all cases the overall behaviour predicted by Simons and
Altshuler was more or less observed but in nearly all cases there had been
significant deviations as well.

This was our motivation to study velocity distributions and correlations in
various mircowave billiards by applying different types of parameter
variations. Most to our surprise we found that the velocity distributions
show a clear dependence on the type of parameter variation. Only for the type
classified as `global' below, we found Simons' and Altshuler's universal
behaviour, whereas for a second type denoted `local' a completely different
behaviour was found. A preliminary version of this paper has been published
already in \cite{bar99b}.

\section{Experiment}

\begin{figure}[t]
\centerline{
\resizebox{6cm}{!}{\includegraphics{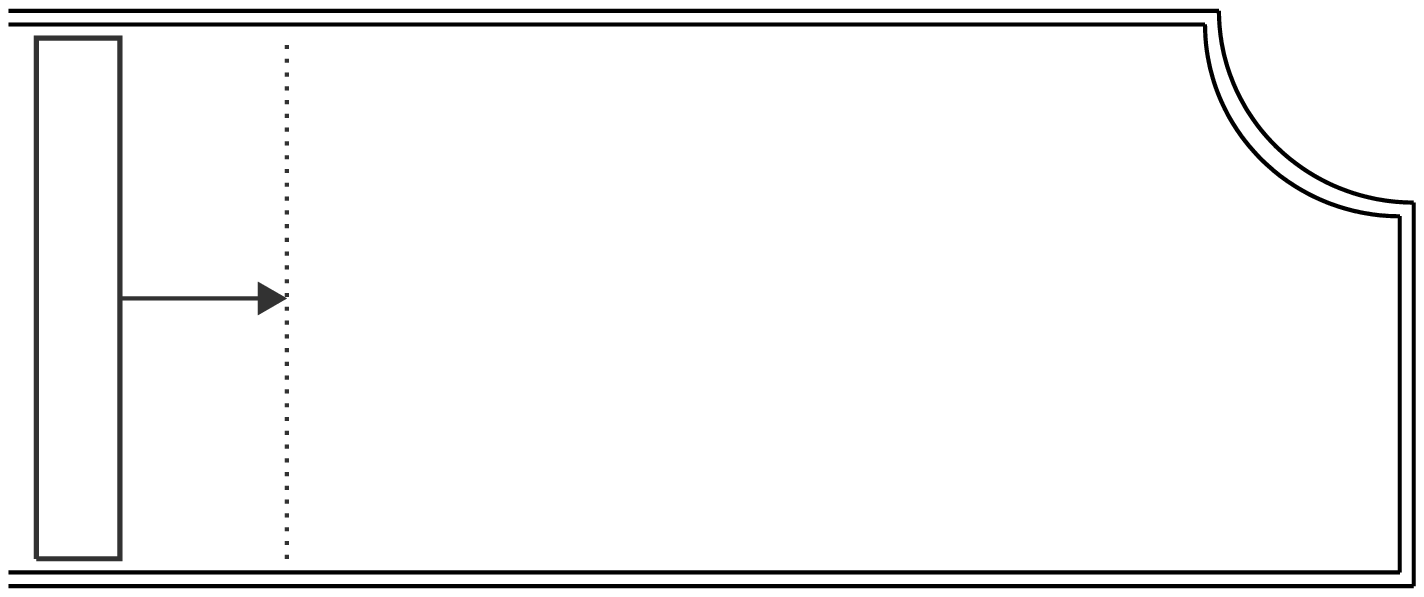}}
\hfill
\resizebox{6cm}{!}{\includegraphics{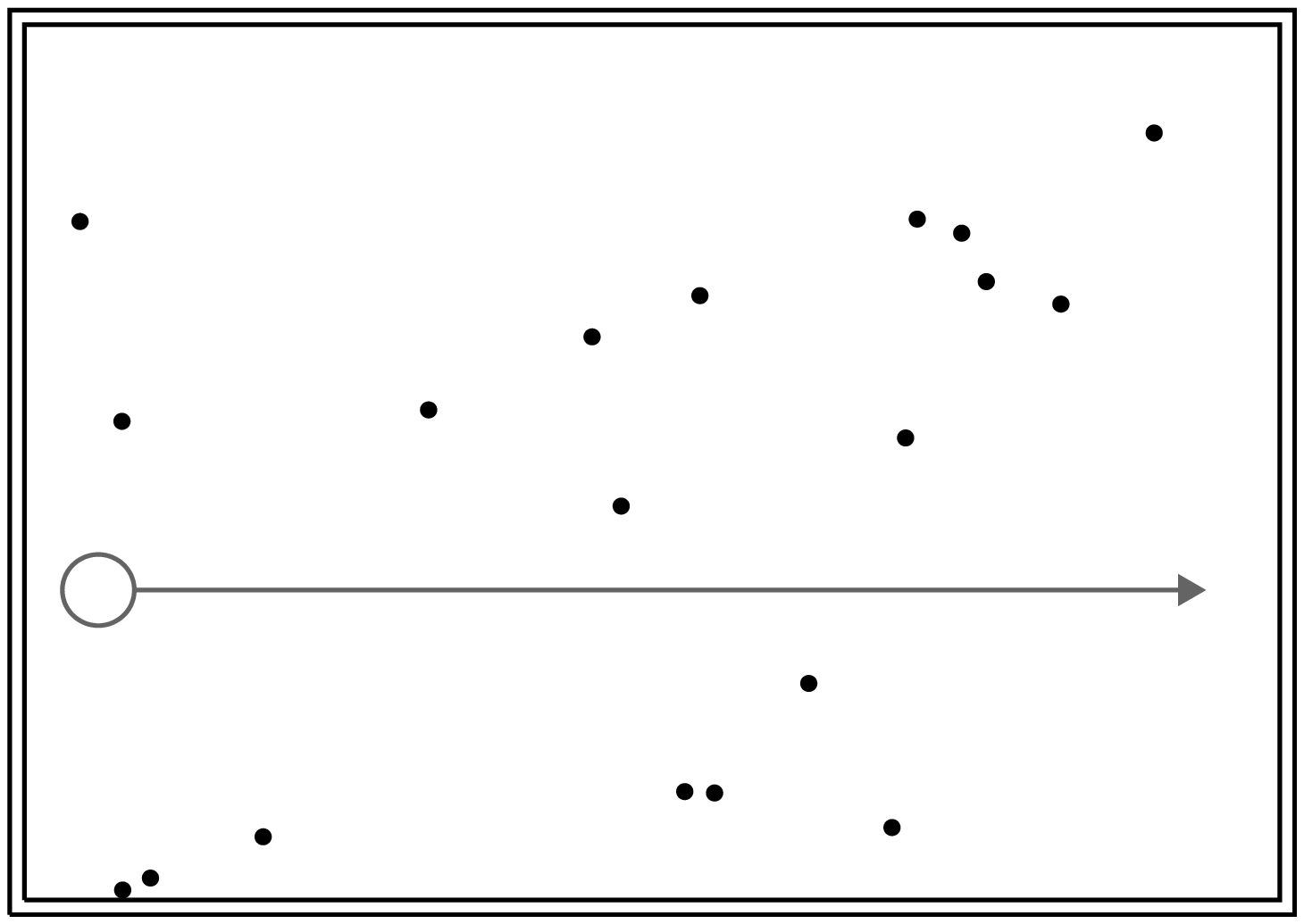}}
}
\caption{Sketch of the quarter Sinai billiard used for the global
level dynamics (left) and of the rectangular billiard with randomly
distributed scatterers (right). Both figures are drawn in scale,
including size and positions of the scatterers. For the dimensions
see the text.}
\label{FigSketch}
\end{figure}

All results presented below have been obtained by taking the eigenfrequency
spectra of billiard-shaped microwave cavities, shortly termed microwave
billiards. The technique is described in \cite{ste95}. Here it may be
sufficient to note that in quasi-two-dimensional resonators, i.e. cavities
with top and bottom plate parallel to each other, quantum mechanical and
electromagnetic spectra are completely equivalent, as long as
the maximum frequency $\nu_{\rm max}=c/2h$ is not surpassed, where
$h$ is the height of the resonator. In the present experiment
the height was 8 mm yielding a maximum frequency of 18.74 GHz.

In the very beginning it was not evident at all which billiard
parameter should be varied. According to Simons and Altshuler
the parameter $X$ `could denote the strength of some field, like
the Aharonov-Bohm flux through a ring, or a magnetic field, or
the position of some impurity' (cited literally from Ref. \cite{sim93b}).
In microwave billiards the magnetic field is not available as a
parameter (though the use of billiards with ferrite-coated walls
may offer an alternative \cite{so95}). On the other hand it is
straightforward to study the spectra of microwave billiards as a
function of some length \cite{kol94a}. The magnetic field and the
length variation have in common that already a small change in the
parameter will modify the wavefunction everywhere.
For this reason we call this type of level dynamics `global'.
The variation of the position of one impurity, on the other hand,
can be easily performed in microwave billiards, with the additional
advantage that, in contrast to real mesoscopic systems, we can
control the disorder in every moment perfectly. The displacement
of one impurity changes the value of the wave function only in its
neighbourhood. Consequently, we call this type of level dynamics `local'.

\begin{figure}[t]
\centerline{
\hspace*{0.4cm}
\resizebox{5.8cm}{!}{\includegraphics{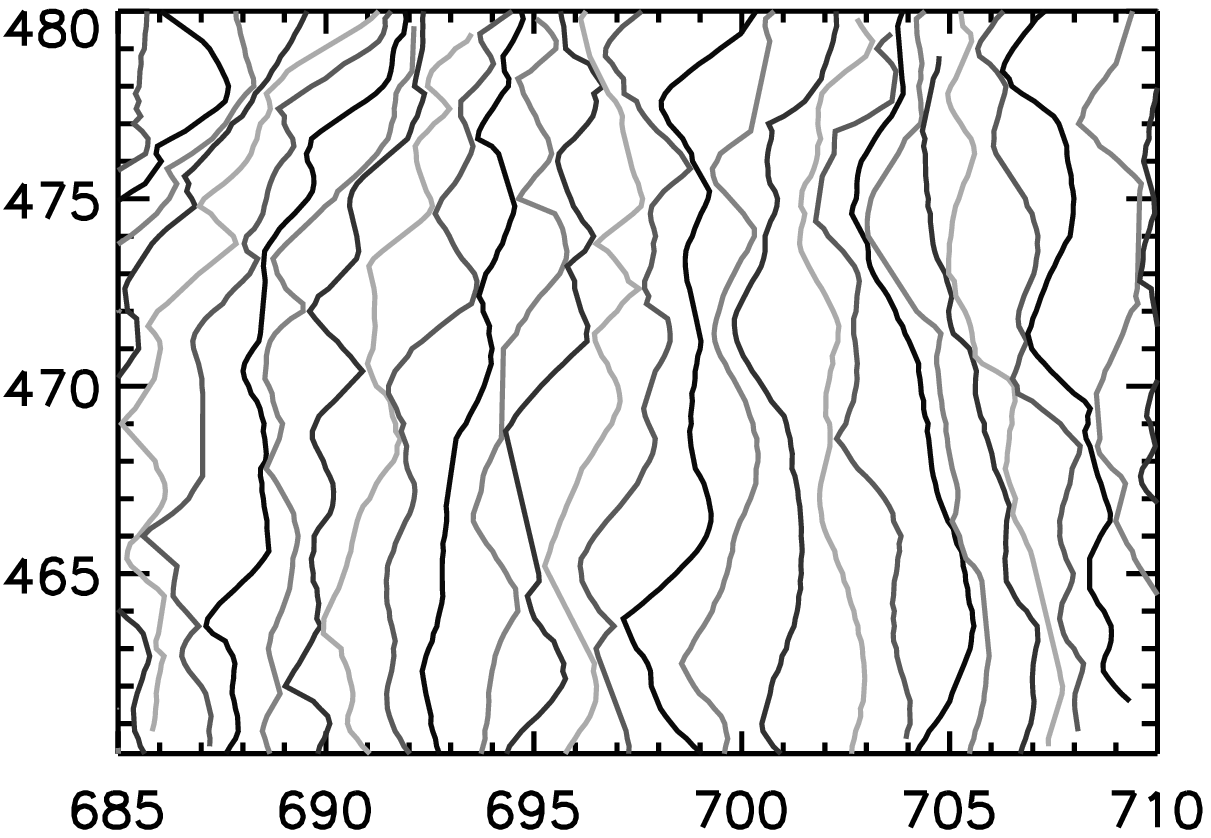}}
\hspace*{0.6cm}
\resizebox{5.8cm}{!}{\includegraphics{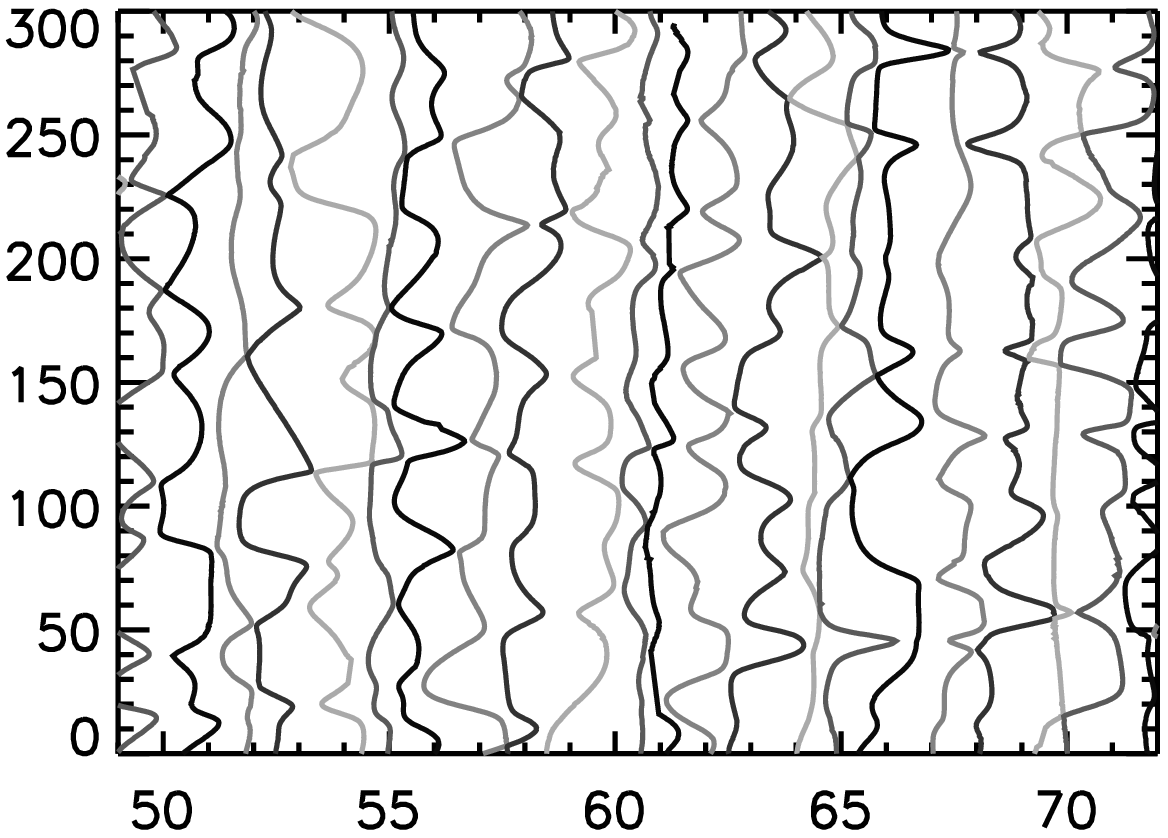}}
}
\vspace*{0.2cm}
\caption{Part of the eigenvalue level dynamics spectra for the quarter
Sinai billiard as a function of the length (left) and the rectangular
billiard with randomly distributed scatterers as a function of the
position of one scatterers (right). All spectra have been unfolded to a
mean density of one.}
\label{FigLeveldyn}
\end{figure}

Both types of level dynamics were realized in the experiments.
One of the systems studied was a quarter Sinai billiard with width
$b$=200 mm, radius $r$=70 mm of the quarter circle, and a length $a$
which was varied between 460 and 480 mm in steps of 0.2 mm. About
120 eigenvalues entered into the data analysis in the frequency
range 14.5 to 15.5 GHz. The second system was a rectangular billiard
with side lengths $a$=340 mm, $b$=240 mm, containing 19 circular disks
with a diameter of 4.6 mm whose positions were determined by means of
a random number generator. The position of one further disk was
varied in one direction in steps of 1 mm. Two different movable disks
were used with diameters of 4.6 and 20 mm, respectively
(see Fig. \ref{FigSketch}).

A part of the level dynamic spectrum of the Sinai billiard is shown in
Fig. \ref{FigLeveldyn} (left). As a function of length the resonances
experience four to five avoided crossing in the studied length range.
This is about the maximum range usable for the velocity correlation
measurements for the following reason. Upon variation of the length it
happens that a node line of an eigenfunction passes the position of the
coupling antenna, leading to a temporary loss of the eigenfrequency. In
the studied length range each resonance disappeared at most once, but
with increasing length range the tieing up of the open ends
became more and more unreliable. The situation changes qualitatively
of the level dynamics of the rectangular billiard with randomly
distributed scatterers shown in Fig. \ref{FigLeveldyn} (right).
Already a comparison by eyes shows that there is much more regularity
in the latter case. Now it is no problem to follow each eigenvalue
over about 10 avoided crossings.

\section{Global level dynamics}

\begin{figure}[t]
\centerline{
\resizebox{6cm}{!}{\includegraphics{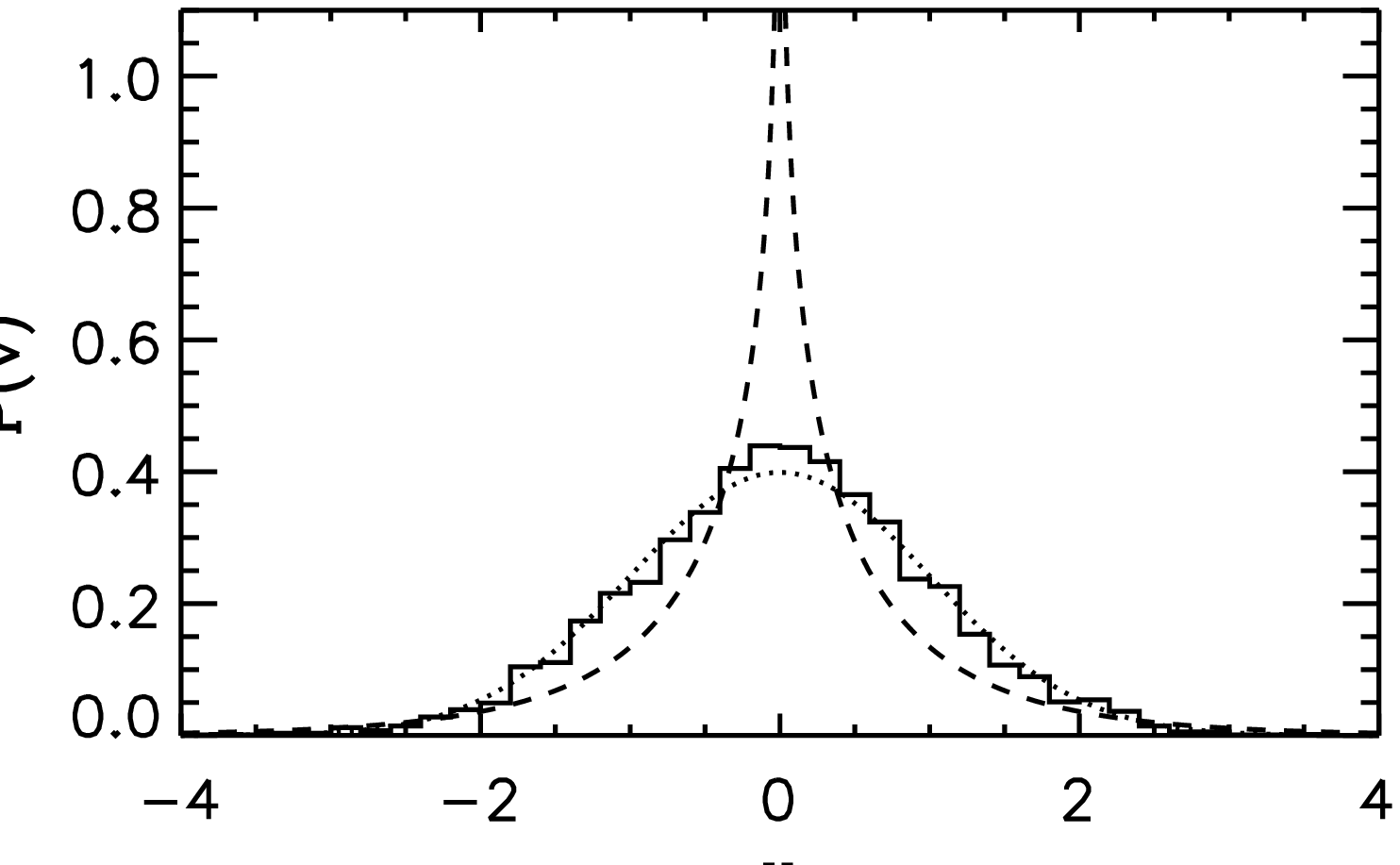}}
\hfill
\resizebox{6cm}{!}{\includegraphics{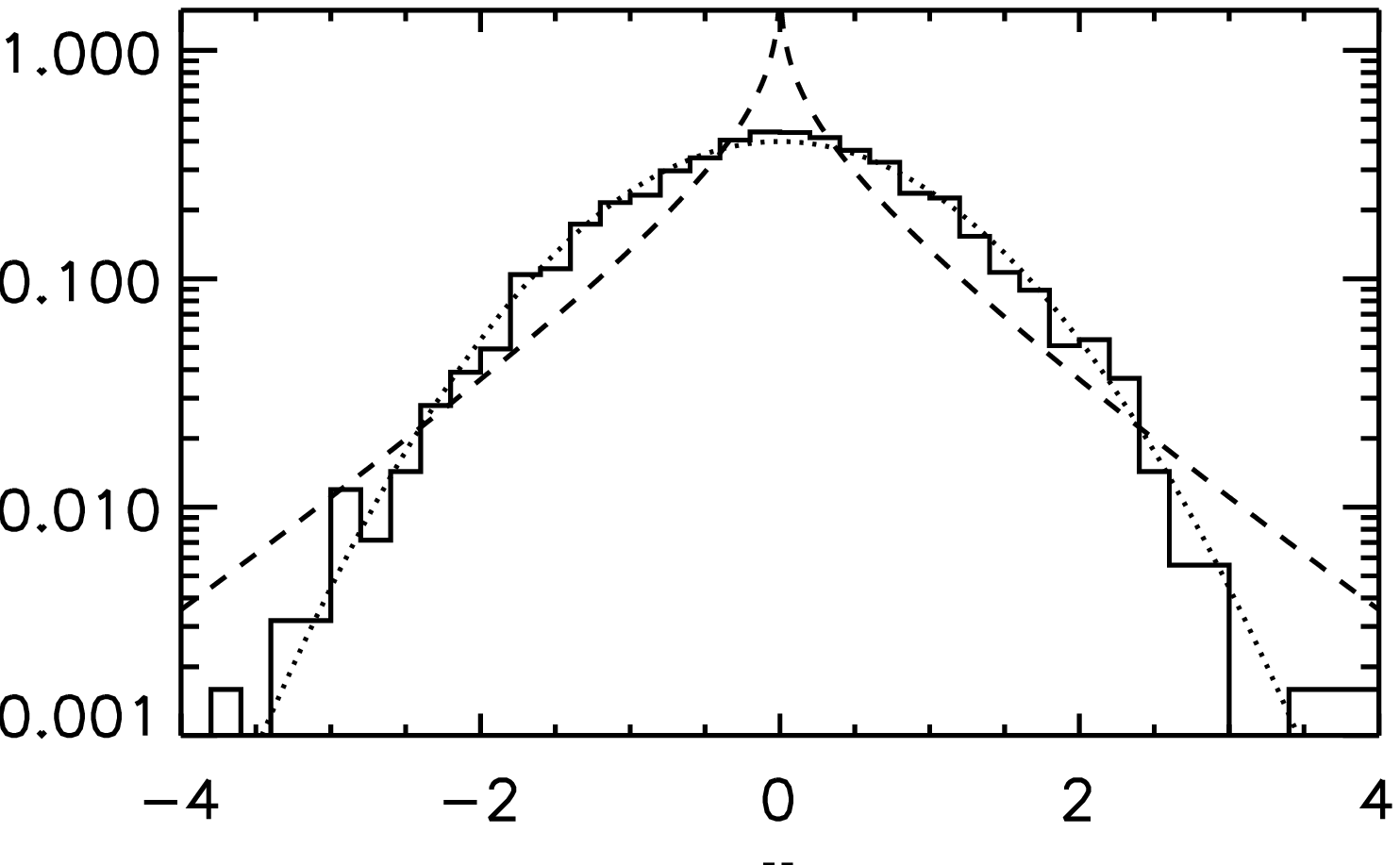}}
}
\caption{Velocity distribution for the quarter Sinai billiard on a linear
(left) and a logarithmic (right) scale. The dotted line corresponds to a
Gaussian, the dashed line to a modified Bessel function (see Eq.
(\ref{EqBessel})).}
\label{FigGlobVel}
\end{figure}

We start with a discussion of the Sinai billiard results. Fig.
\ref{FigGlobVel} shows the found velocity distribution both on a linear
and a logarithmic scale. The found distribution is close to a Gaussian
in accordance with the universal prediction. The dashed line is a
modified Bessel function which will become important in the following.
It is well-known that non-generic features such as bouncing balls give
rise to deviations from a Gaussian distribution in particular in the
region of the wings \cite{kol94a,sie95}. In the measurement we therefore
carefully avoided the regions disturbed by the dominating bouncing ball.
Hence such a disturbance does not show up in the present case.

\begin{figure}[t]
\centerline{\resizebox{6cm}{!}{\includegraphics{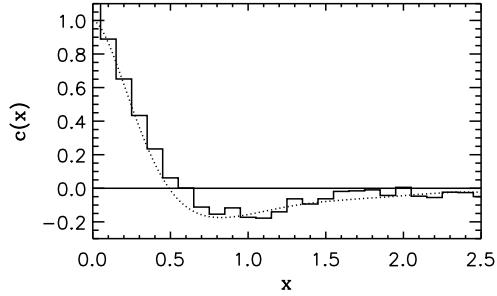}}}
\caption{Velocity autocorrelation function for the quarter Sinai billiard
as a function of the position of one length rescaled according to Eq.
(\ref{EqScale}). The dotted line corresponds to Simons' and Altshuler's
universal function.}
\label{FigGlobAC}
\end{figure}

Fig. \ref{FigGlobAC} shows the corresponding velocity autocorrelation
function. To obtain the result, each eigenvalue was followed over the
available length range including four to five avoided crossings. The mean
squared velocity needed for the scaling (\ref{EqScale}) was calculated for
each eigenvalue independently. Subsequently the results of all eigenvalues
were superimposed. Such a procedure was mandatory, since the individual
mean squared velocities showed large variations.
For this reason a scaling with a globally calculated mean squared velocity
did not give satisfactory results.

The dotted line in Fig. \ref{FigGlobAC} corresponds to Simons' and
Altshuler's universal velocity correlation function. Since an analytical
expression for this quantity is not available, it was obtained from a random
matrix simulation \cite{muc}. The overall agreement between experiment and
theory is good, apart from some small but significant deviations close to the
minimum. It was already mentioned in the beginning that in most studied cases
such an agreement was not found.

\section{Local level dynamics}

\begin{figure}[t]
\centerline{
\resizebox{6cm}{!}{\includegraphics{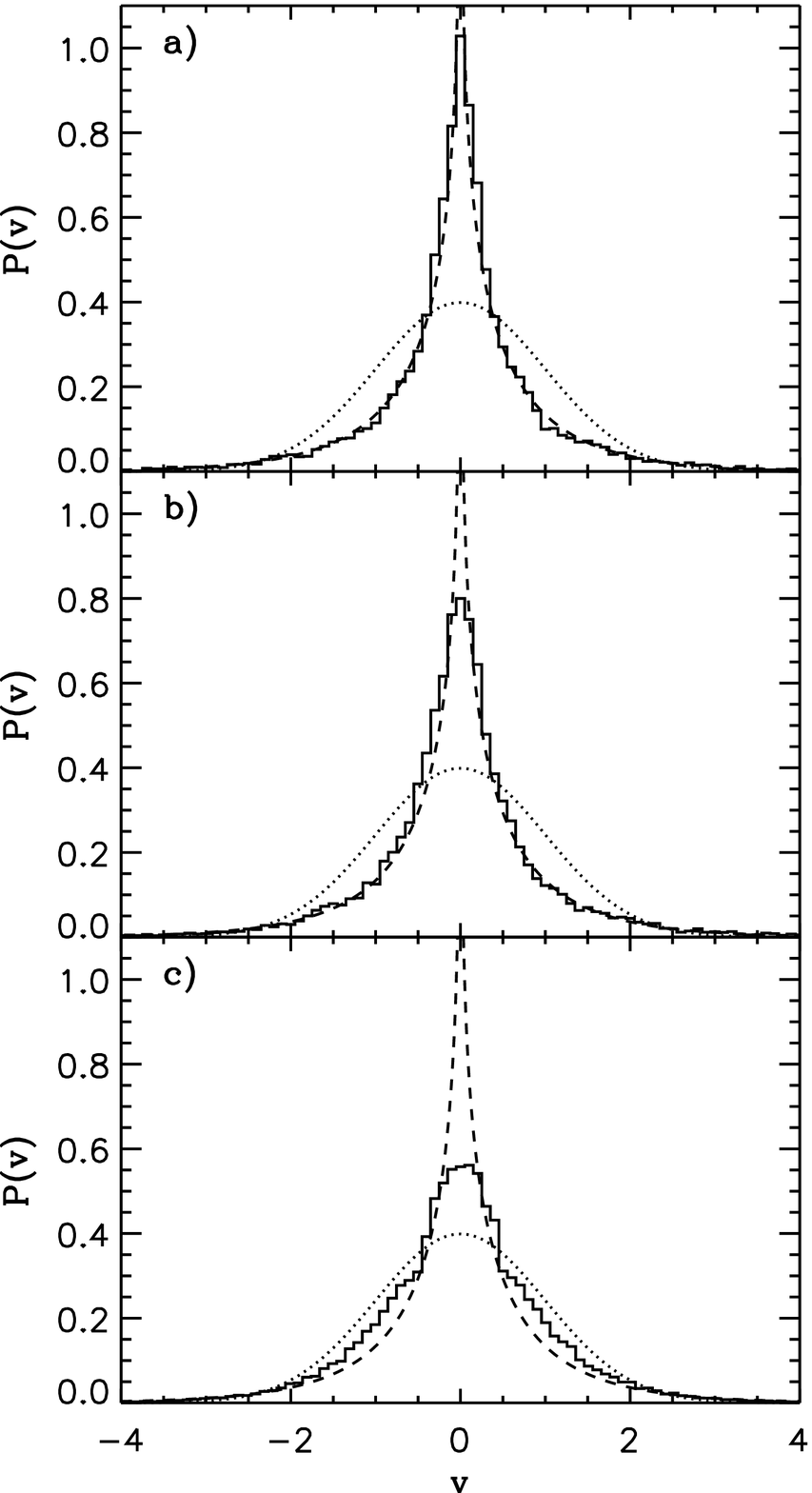}}
\hfill
\resizebox{6cm}{!}{\includegraphics{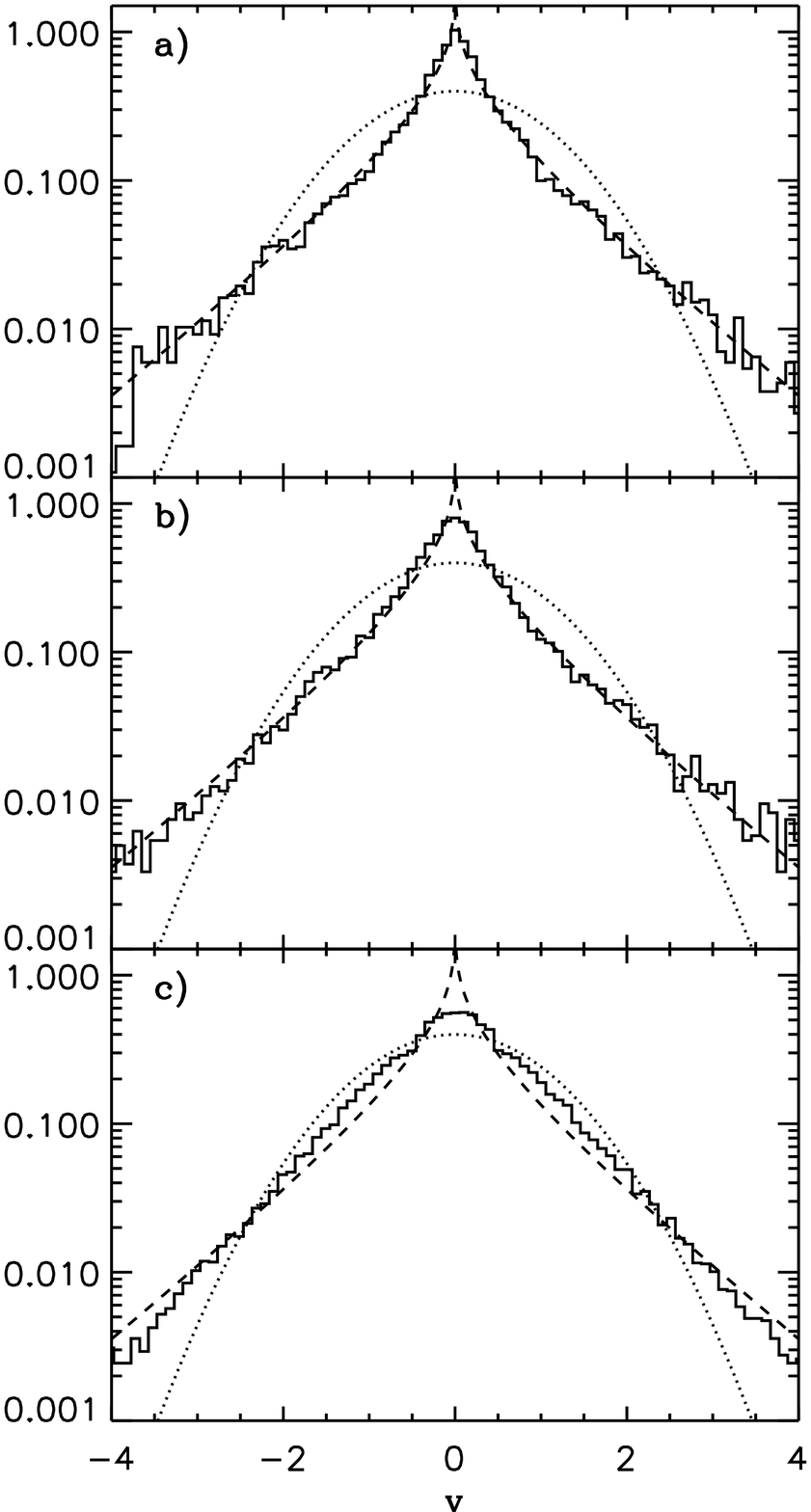}}
}
\caption{Velocity distributions in a rectangular billiard with randomly
distributed scatterers as a function of the position of one of the
scatterers on a linear (left column) and a logarithmic (right column) scale.
The histograms were taken in three different ranges of $\delta=kD$, where $D$
is the diameter of the movable scatterer: $0.35<\delta<0.65$ (a),
$1.4<\delta<2.6$ (b), $5.1<\delta<5.9$ (c). The dotted line corresponds to a
Gaussian, the dashed line to a modified Bessel function.}
\label{FigLocVel}
\end{figure}

\begin{figure}[t]
\centerline{
\resizebox{6cm}{!}{\includegraphics{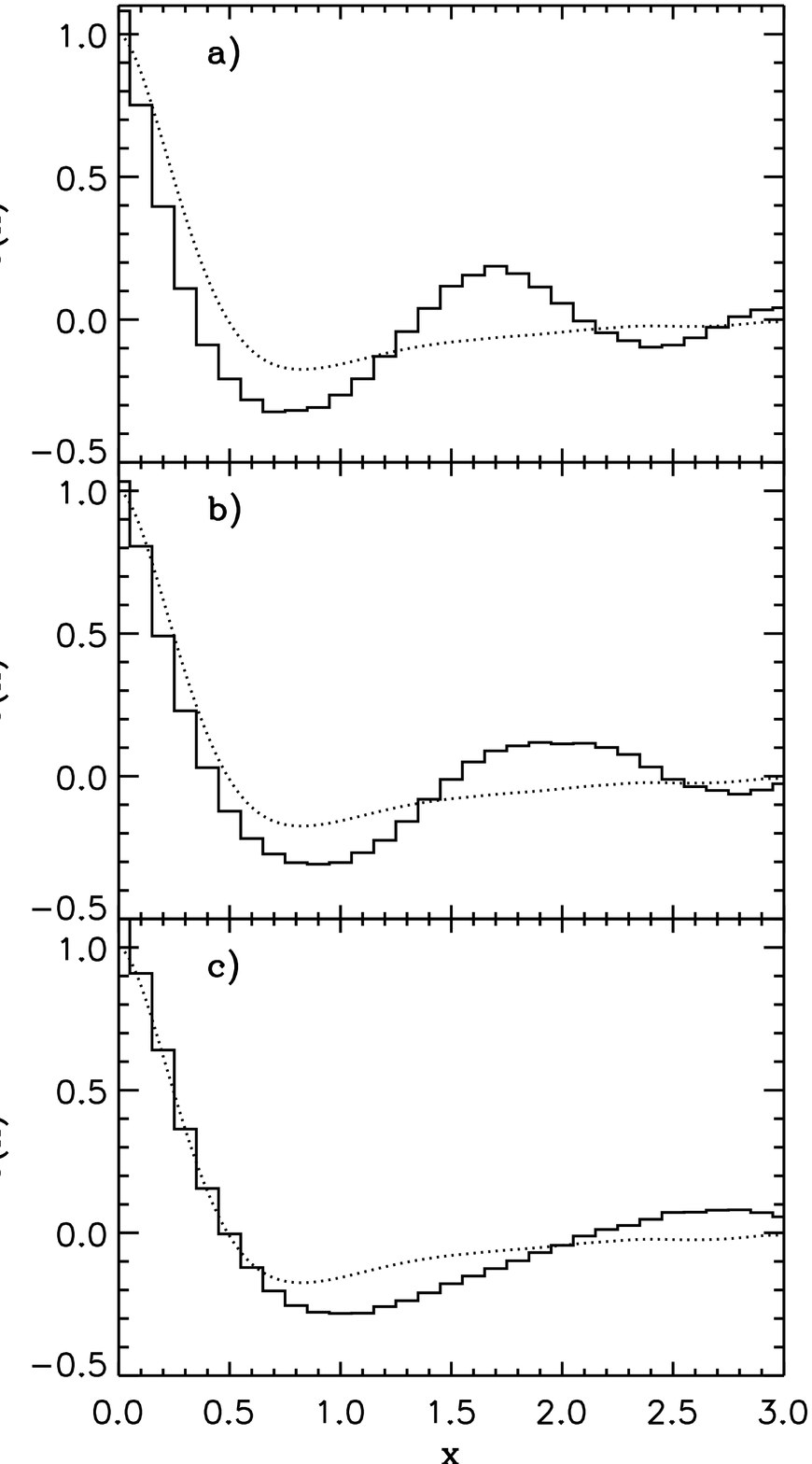}}
\hfill
\resizebox{6cm}{!}{\includegraphics{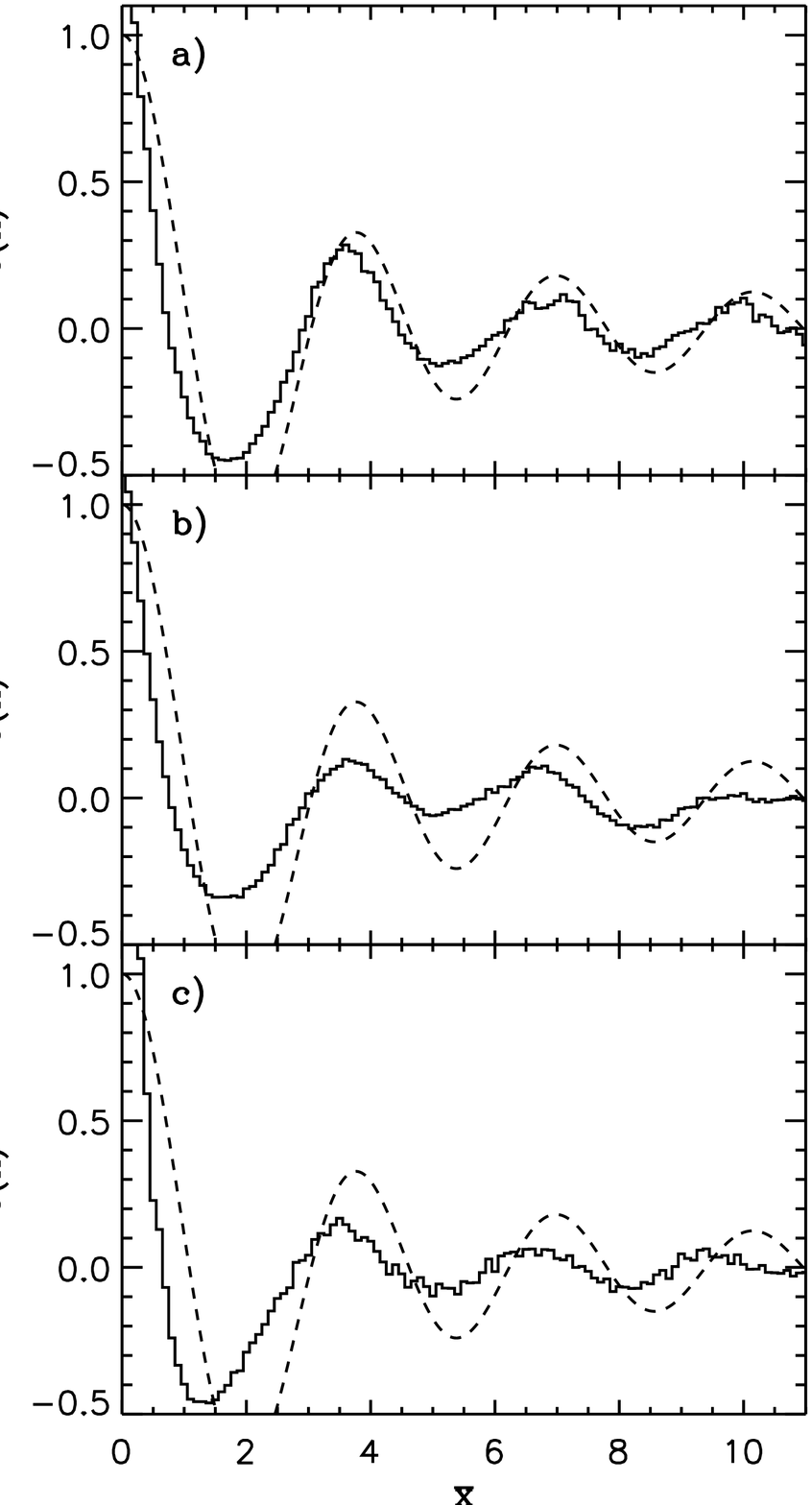}}
}
\caption{Velocity autocorrelation functions in a rectangular billiard with
randomly distributed scatterers for the same ranges of $\delta$ as in Fig.
\ref{FigLocVel}. On the left column the position of the movable scatterer was
scaled according to Eq. (\ref{EqScale}). The dotted line is Simons' and
Altshuler's universal function. On the right column the same data are plotted,
but with a scaling of the position according to Eq. (\ref{EqNewX}). The dashed
lines have been calculated from Eq. (\ref{EqLocalAC}).}
\label{FigLocAC}
\end{figure}

Whether a movable scatterer perturbs the system globally or locally, depends
on its size, or, more precisely, on the parameter $\delta=kD$, where $D$ is
the diameter of the scatterer and $k$ the wavenumber. For small $\delta$
values ($\delta \ll 1$) a local level dynamics is expected, whereas with
increasing $\delta$ values a gradual transition towards a global behaviour
should be observed. Fig. \ref{FigLocVel} shows the velocity distribution in
the rectangular billiard with randomly distributed scatterers for three
different $\delta$ ranges, again in a linear and a logarithmic plot. In Fig.
\ref{FigLocVel}(a) the diameter of the movable disk was 4.6 mm, and the
eigenvalues were taken in the range 3.4 to 6 GHz. In Figs. \ref{FigLocVel}(b)
and (c) the diameter of the scatterer was 20 mm, and the eigenvalues were in
the range 3.4 to 5.8 GHz and 12.5 to 14.3 GHz, respectively.

For small $\delta$ values the found velocity distribution is well described by
a modified Bessel function
\begin{equation}
P(v) = \frac{1}{\pi\sqrt{\left<v^2\right>}}
K_0\left(\frac{|v|}{\sqrt{\left<v^2\right>}}\right).
\label{EqBessel}
\end{equation}

The logarithmic plot shows that this behaviour is found over two orders of
magnitude. Note that now asymptotically a single exponential behaviour is
found in contrast to the Gaussian behaviour observed for the global
velocities.

With increasing $\delta$ values a transition to Gaussian behaviour is
observed as expected. We completed the sequence with a level dynamics in a
Sinai billiard, where the central circle was moved, corresponding to $\delta$
values between 30 and 40. Here the found velocities were again Gaussian
distributed (not shown).

A corresponding deviation from the universal behaviour is found for the
velocity autocorrelation functions. The left column of Fig. \ref{FigLocAC}
shows the results for the same $\delta$ ranges as in Fig. \ref{FigLocVel}.
There is no similarity at all with the universal behaviour. Only for the
largest $\delta$ range the found correlation function seems to approach the
universal function.

All these findings are in striking contrast to all what has been expected by
the experts. Nevertheless, the discrepancy from the universal behaviour can
be quite easily resolved. It is a well-known fact that the insertion of a
metallic  object into a quasi-two-dimensional microwave cavity shifts the
eigenfrequencies to smaller frequencies. If the dimensions of the perturber
are small compared to the wavelength, this shift is proportional to the
square of the electric field. This technique has been used routinely for many
years to map field distributions in microwave cavities \cite{mai52a} and has
recently been applied to the study of wave function in chaotic billiards as
well \cite{sri91,doer98b,wu98}. Applied to the present problem this means
that in the limit $\delta\to 0$ the eigenvalue velocities are given by
\begin{equation}
\frac{\partial E_n}{\partial X} = \alpha \nabla|\psi|^2
\label{EqVelocity}
\end{equation}
where $X$ is the position of the scatterer, and $\nabla$ is the gradient
taken in the direction of the displacement. $\alpha$ is a constant factor
depending exclusively on the geometry of the perturber. It follows for the
velocity distribution
\begin{equation}
P(v) = \left< \delta(v-2\alpha\psi\nabla\psi) \right>
\label{EqvAverage}
\end{equation}
where the brackets denote an average of eigenvalues. This average can be
calculated, if we assume that the wavefunction $\psi(r)$ can be described by
a random superposition of plane waves \cite{ber77a}
\begin{equation}
\psi(r) = \sum a_n e^{i \vec k_n \vec r}
\label{EqPlaneWaves}
\end{equation}
where the modulus of the wavenumbers of all waves contributing is the
same, $|\vec k_n| = k$, but where amplitudes $a_n$ and directions
$\vec k_n / k$ are random. This approach has been proved extremely
fruitful in the description of eigenfunctions in chaotic systems
\cite{ocon87,mcd88,doer98b}.

By means of the central limit theorem it is now an easy matter to show that
under these assmptions $P(\psi)$ and $P(\nabla\psi)$ are uncorrelated and both
Gaussian distributed. With these ingredients the average (\ref{EqvAverage})
can be calculated, resulting in the modified Bessel function (\ref{EqBessel})
for the velocity distribution. With help of the ansatz (\ref{EqPlaneWaves}) the
velocity autocorrelation can be calculated as well, using standard techniques
\cite{sre96b}. The result is
\begin{equation}
c(\bar{x}) = - \left[ J_0^2(\bar{x}) \right]''
= J_0^2(\bar{x}) - 2J_1^2(\bar{x}) - J_0(\bar{x}) J_2(\bar{x})
\label{EqLocalAC}
\end{equation}
where
\begin{equation}
\bar{x} = k X.
\label{EqNewX}
\end{equation}

Note that in contrast to Simons' and Altshuler's approach where the level
dynamics parameter $X$ has been scaled by means of the square root of the
quadratically averaged velocities, now we have got a straightforward scaling
in terms of the wavenumber $k$.

On the right column of Fig. \ref{FigLocAC} the same data shown on the
left column are plotted again, but now with an abscissa scaled according
to Eq. (\ref{EqNewX}). The solid line has been calculated from Eq.
(\ref{EqLocalAC}). It describes the experimental results quite well. In
particular the wavelengths of the oscillations are reproduced correctly even
for the larger $\delta$ ranges. Only an increasing damping of the oscillations
with increasing $\delta$ is observed. This oscillatory behaviour in the
velocity autocorrelation function, which is already evident from a visual
inspection of Fig. \ref{FigLeveldyn} (right), is in sharp contrast to the
global case, where the autocorrelation function shows only one minimum,
while all subsequent oscillations found for the individual levels are
completely wiped out by averaging process.

\section{Conclusion}

This work has shown that in contrast to our previous understanding there is
not one but there are two regimes of universality. Only for the global case
where a small variation of the level dynamics parameter changes the wave
functions everywhere, the universal velocity distributions and correlations
predicted by Simons and Altshuler are really observed. In this range the
scaling of the level dynamics parameter in terms of the square root of the
quadratically averaged velocities is the appropriate one.

In the local limit, on the other hand, realized in the present work by
moving a small pertuber through a disordered system, another universal
regime is observed. Here the scaling of the level dynamics parameter in
terms of the wave number is the correct one, and another class of velocity
distributions and correlations is found, which has been discussed in this
paper.

\vspace*{0.25cm}
\baselineskip=10pt{\small \noindent
The issue of this paper has been discussed with numerous colleagues from
the mesoscopic community. Discussions with Y. Fyodorov and T. Guhr have
been in particular helpful. E. Mucciolo provided us with his unpublished
calculation of the universal velocity autocorrelation function. The work
has been supported by the Deutsche Forschungsgemeinschaft via the
SFB 185 "Nichtlineare Dynamik" as well as by an individual grant.}


\begin{thebibliography}{10}
\bibitem{edw72}
J.T. Edwards and D.J. Thouless, J. Phys. C {\bf 5} (1972) 807
\bibitem{tho74}
D.J. Thouless, Phys. Rep. {\bf 13} (1974) 93
\bibitem{akk92}
E. Akkermans and G. Montambaux, Phys. Rev. Lett. {\bf 68} (1992) 642
\bibitem{sza93}
A. Szafer and B.L. Altshuler, Phys. Rev. Lett. {\bf 70} (1993) 587
\bibitem{sim93b}
B.D. Simons and B.L. Altshuler, Phys. Rev. B {\bf 48} (1993) 5422
\bibitem{sim95}
B.L. Altshuler and B.D. Simons, in {\it Mesoscopic Quantum Physics}
edited by E. Akkermans {\it et~al.}, North Holland, Amsterdam 1995, p. 1
\bibitem{sim93c}
B.D. Simons {\it et~al.}, Phys. Rev. Lett. {\bf 71} (1993) 2899
\bibitem{bru96}
H. Bruus, C.H. Lewenkopf, and E.R. Mucciolo,
Phys. Rev. B {\bf 53} (1996) 9968
\bibitem{hlu}
Y. Hlushchuk {\it et~al.}, to be published.
\bibitem{ber}
P. Bertelsen {\it et~al.}, Phys. Rev. Lett. {\bf 83} (1999) 2171
\bibitem{bar99b}
M. Barth, U. Kuhl, and H.-J. St\"ockmann,
Phys. Rev. Lett {\bf 82} (1999) 2026
\bibitem{ste95}
J. Stein, H.-J. St\"ockmann, and U. Stoffregen,
Phys. Rev. Lett. {\bf 75} (1995) 53
\bibitem{so95}
P.~So {\it et~al.}, Phys. Rev. Lett. {\bf 74} (1995) 2662
\bibitem{kol94a}
M. Kollmann {\it et~al.}, Phys. Rev. E {\bf 49} (1994) R1
\bibitem{sie95}
M. Sieber {\it et~al.}, J. Phys. A {\bf 28} (1995) 5041
\bibitem{muc}
E.R. Mucciolo, private communication
\bibitem{mai52a}
L.C. Maier and J.C. Slater, J. Appl. Phys. {\bf 23} (1952) 78
\bibitem{sri91}
S. Sridhar, Phys. Rev. Lett. {\bf 67} (1991) 785
\bibitem{doer98b}
U. D\"orr {\it et~al.}, Phys. Rev. Lett. {\bf 80} (1998) 1030
\bibitem{wu98}
D.H. Wu {\it et~al.}, Phys. Rev. Lett. {\bf 81} (1998) 2890
\bibitem{ber77a}
M.V. Berry, J.~Phys.~A {\bf 10} (1977) 2083
\bibitem{ocon87}
P. O'Connor, J. Gehlen, and E.J. Heller,
Phys. Rev. Lett. {\bf 58} (1987) 1296
\bibitem{mcd88}
S.W. McDonald and A.M. Kaufman, Phys. Rev. A {\bf 37} (1988) 3067
\bibitem{sre96b}
M. Srednicki and F. Stiernelof, J. Phys. A {\bf 29} (1996) 5817
\end{thebibliography}
\end{document}